\documentclass[preprint,journal]{vgtc}       

\ifpdf
  \pdfoutput=1\relax                   
  \usepackage{graphicx}                
  \DeclareGraphicsExtensions{.pdf,.png,.jpg,.jpeg} 
\else
  \usepackage{graphicx}                
  \DeclareGraphicsExtensions{.eps}     
\fi%

\graphicspath{{figures/}{pictures/}{images/}{./}} 

\usepackage{xcolor}
\definecolor{mygreen}{HTML}{9cb0a2}
\definecolor{mybrown}{HTML}{c39b83}
\definecolor{mypurple}{HTML}{bbafd1}

\usepackage{comment}

\usepackage{multirow}

\usepackage{paralist} 

\usepackage{float}

\usepackage{algorithm}
\usepackage{algpseudocode}

\usepackage{amsmath}

\usepackage{bbm}

\usepackage{url}

\usepackage{enumitem}

\newcommand*{\doi}[1]{\href{http://dx.doi.org/#1}{doi: #1}}

\usepackage{microtype}                 
\PassOptionsToPackage{warn}{textcomp}  
\usepackage{textcomp}                  
\usepackage{mathptmx}                  
\usepackage{times}                     
\usepackage{cite}                      
\usepackage{tabu}                      
\usepackage{booktabs}                  

\onlineid{1478}

\vgtccategory{Research}
\vgtcpapertype{Analytics \& Decisions}

\title{DPVisCreator: Incorporating Pattern Constraints to Privacy-preserving Visualizations via Differential Privacy}

\author{Jiehui Zhou, Xumeng Wang, Jason K. Wong, Huanliang Wang, Zhongwei Wang, Xiaoyu Yang, \\ Xiaoran Yan, Haozhe Feng, Huamin Qu, Haochao Ying, and Wei Chen}

\authorfooter{
\item
J. Zhou, H. Wang, Z. Wang, X. Yang, H. Feng, and W. Chen are with the State Key Lab of CAD\&CG, Zhejiang University. W. Chen is also with the Laboratory of Art and Archaeology Image (Zhejiang University), Ministry of Education, China. E-mail: \{zhoujiehui, 22051090, wzw09, 22051142, fenghz, chenvis\}@zju.edu.cn.

\item
X. Wang is with TMCC, CS, Nankai University. E-mail: wangxumeng@nankai.edu.cn.

\item
J. Wong and H. Qu are with the Hong Kong University of Science and Technology. Email: \{kkwongar, huamin\}@cse.ust.hk.

\item
X. Yan is with the Zhejiang Lab. E-mail: xiaoran.a.yan@gmail.com.

\item
H. Ying is with the School of Public Health, Zhejiang University. He is also with the Key Laboratory of Intelligent Preventive Medicine of Zhejiang Province. E-mail: haochaoying@zju.edu.cn.

\item
Haochao Ying and Wei Chen are the corresponding authors.
}

\shortauthortitle{Biv \MakeLowercase{\textit{et al.}}: Global Illumination for Fun and Profit}

\abstract{
Data privacy is an essential issue in publishing data visualizations. However, it is challenging to represent multiple data patterns in privacy-preserving visualizations. The prior approaches target specific chart types or perform an anonymization model uniformly without considering the importance of data patterns in visualizations. In this paper, we propose a visual analytics approach that facilitates data custodians to generate multiple private charts while maintaining user-preferred patterns. To this end, we introduce pattern constraints to model users' preferences over data patterns in the dataset and incorporate them into the proposed Bayesian network-based Differential Privacy (DP) model \textit{PriVis}. A prototype system, \textit{DPVisCreator}, is developed to assist data custodians in implementing our approach. The effectiveness of our approach is demonstrated with quantitative evaluation of pattern utility under the different levels of privacy protection, case studies, and semi-structured expert interviews.

} 


\CCScatlist{ 
 \CCScat{K.6.1}{Management of Computing and Information Systems}%
{Project and People Management}{Life Cycle};
 \CCScat{K.7.m}{The Computing Profession}{Miscellaneous}{Ethics}
}

\teaser{
  \centering
  \includegraphics[height=7cm, keepaspectratio]{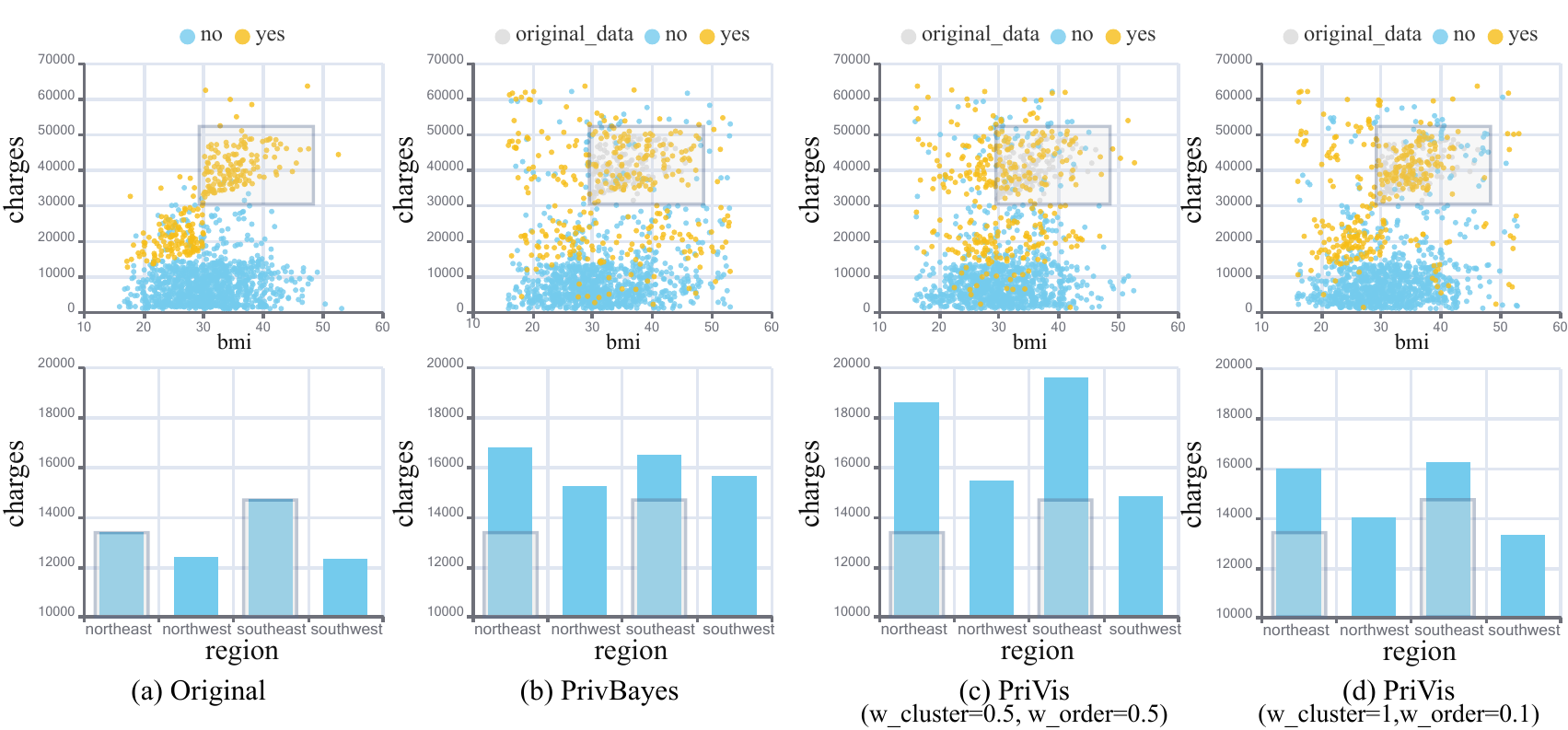}
  \caption{Comparison of a series of privacy-preserving visualization charts under the same privacy budget. (a) The original data. (b) The PrivBayes~\cite{zhang2017privbayes} results, where the cluster in the gray boxes are barely retained, and the ranking of the selected bars changes. (c, d) The \textit{PriVis} results with different pattern constraints, showing that adjusting the importance weights can help maintain different patterns.}
  \label{fig:teaser}
}
\newcommand{\rc}[1]{\textcolor{black}{#1}}
\newcommand{\rw}[1]{\textcolor{black}{#1}}

\vgtcinsertpkg

\begin{document}
	


\firstsection{Introduction}

\maketitle


Data visualization is widely applied to support the exploration and understanding of domain data patterns~\cite{healy2014data, o2010visualizing, o2018visualization}. For instance, visualizations help medical practitioners analyze the development of patients' diseases by intuitively showing their health status throughout time~\cite{perer2012matrixflow}. However, when the analyzed data contains sensitive information, unrestricted and unscrutinized representations of such data can result in privacy leakage. A large number of data breaches~\cite{trautman2016corporate, steel2010facebook, craig2017post} have caused severe financial losses and reputation crises. Disclosing sensitive disease information may also provoke discrimination against patients and carry legal consequences~\cite{price2019privacy}.
Many countries and jurisdictions have tightened privacy regulations to protect their citizens' data security and privacy rights. For example, the EU enacted the GDPR to strengthen individuals' control over personal data. In the UK, health data must be processed and analyzed in the Trusted Research Environment (TRE) to ensure that no personal information is revealed in any analytical results. 
Therefore, visualization publishing has an urgent requirement of privacy preservation. 

However, it is challenging to generate a group of privacy-preserving visualizations that represent multiple data patterns to support comprehensive data understanding. 
Privacy-preserving visualizations can be achieved through screen-space sanitization and data-space sanitization~\cite{dasgupta2011adaptive}. Screen-space sanitization creates visual clutter by overlapping~\cite{dasgupta2012conceptualizing} and splitting~\cite{dasgupta2013measuring} elements to obfuscate the visual effects of sensitive information. However, privacy preservation from the perception perspective is disputable without extensive evaluation. Existing data-space sanitization~\cite{chou2017privacy, chou2019privacy, zhang2016challenges} applies automatic models to detect privacy issues and handle them by manipulating the data directly. Recent advanced techniques such as Differential Privacy (DP)~\cite{dwork2006calibrating} are theoretically proven to avoid privacy disclosure. However, these approaches are not designed for visualization; They perform privacy-preserving processing at each data point uniformly without considering the importance of the data patterns when visualized. Therefore, important patterns may be highly distorted or even destroyed. Visualizing these results would decrease the usability of the charts and increase the risk of communicating incorrect information. 


To address this challenge, we develop a novel visual analytics approach that enables data custodians to generate privacy-preserving visualizations that reserve preferred patterns. We introduce \textit{pattern constraints} to model users' preferences over different patterns in a sensitive dataset. The multiple pattern constraints can be represented as a distribution of importance over the entire dataset. We further propose \textit{PriVis}, a privacy-preserving visualization generation model, which extends the Bayesian network-based DP data generation method to consider pattern constraints in the process of structure learning on the data. We present \textit{DPVisCreator}, an interactive prototype system that allows users to interactively explore data patterns in sensitive datasets and specify the pattern constraints. It has a set of coordinated views to help users understand the impacts of different pattern constraints on the privacy-preserving model and the final generated visualizations. It supports efficient privacy-preserving process configuration and intuitive validation of results through user-friendly interactions. \textit{DPVisCreator} is capable of meeting data custodians' needs to publish different types of charts that preserve both privacy and desired patterns. 
Our contributions are as follows:

\begin{itemize}[noitemsep,topsep=0pt]

  \item A privacy-preserving visualization generation model, \textit{PriVis}, to maintain multiple data patterns by extending a DP data publishing model based on Bayesian networks with pattern constraints.

  \item An interactive prototype system, \textit{DPVisCreator}, to specify pattern constraints, understand the privacy-preserving process, and compare various privacy protection schemes.
 
\end{itemize}

\section{\rc{Background}}\label{sec:bac}

\rc{We present an example scenario to illustrate the motivation and application of privacy-preserving visualization. Then, we introduce the basic concepts of DP used in this paper. Finally, we describe evaluation metrics for the utility of the visualization.}

\subsection{\rc{Example Scenario in Healthcare}}

\label{sec:exa}
\rc{Hospitals hold many electronic health records (EHRs) containing patients' private information, such as those in~\autoref{tab:example_data}.}
\rc{In this scenario, there are three stakeholders: \textit{data owner}, \textit{data custodian}, and \textit{data consumer}. The data owner is the hospital, which owns the patients' EHRs. The data custodian is a clinical data analyst with access to the raw data and is tasked with publishing data through visualizations. The data custodian needs to have some basic knowledge of privacy. The data consumer receives the published visualizations for information.
}

\begin{table}[htb]
  \centering
  \caption{An example sensitive EHRs dataset.}
  \label{tab:example_data}
  \begin{tabular}{p{0.3\linewidth} c c c }
  \toprule
  \textbf{Name}  & \textbf{Age} & \textbf{Cholesterol} & \textbf{Heart attack}  \\ 
  \midrule
      Liam Clark & 23 & 6.2 & Yes \\
      Walter Woods & 52 & 2.5 & No \\
      Rebecca Fraser & 30 & 3.8 & No \\
      Caitlyn Gonzales & 42 & 4.2 & Yes \\
      Laura Hernandez & 69 & 3.1 & Yes \\
  \bottomrule
  \end{tabular}
\end{table}

\rc{The data custodian wants to demonstrate that people over 60 are at a much greater risk of heart disease than others. A bar chart can visually summarize the number of heart attack patients in each age group, but it may also disclose private information about individuals' illnesses. For example, if malicious attackers know that Liam is the only patient in the 20-25 age range who had been to the hospital, they could infer that Liam had a heart attack based on the distribution pattern in the bar chart. Leaking such personal information may lead Liam to suffer from targeted advertisements and high premiums for health insurance.}
\rc{Therefore, data custodians face a dilemma between publishing significant data facts and protecting patients' privacy. On the one hand, it is necessary to ensure that the charts do not reveal private information about individuals. On the other hand, privacy-preserving charts should be able to preserve important data patterns and communicate insights to data consumers.
}

\subsection{\rc{Preliminary Privacy Preservation}}


\rc{We focus on the widely used \textit{tabular data}, which stores data records with a set of attributes as rows in a multidimensional table. Privacy preservation for tabular data is mainly achieved by syntactic anonymity and DP. Syntactic anonymity (e.g., $k$-anonymity~\cite{sweeney2002k}, $l$-diversity~\cite{machanavajjhala2007diversity}, and $t$-closeness~\cite{li2007t}) constructs equivalence groups to prevent attackers from distinguishing individuals. Unfortunately, equivalence groups could be cracked by background knowledge~\cite{clifton2013syntactic}.}

\rc{Compared with syntactic anonymity approaches, DP is gradually being applied to more real-world scenarios because of its robust mathematical guarantees~\cite{dwork2006calibrating}.
DP is defined based on two neighboring datasets $D_1$ and $D_2$, which differ by adding or removing a record. Given a user-defined privacy budget $\epsilon$ ($\epsilon>0$), a randomized algorithm $G$ satisfies \textbf{$\epsilon$-differential privacy} ($\epsilon$-DP), if and only if the following equation holds for any possible output $O$:}
\begin{equation}
\operatorname{Pr}\left(G\left(D_{1}\right)=O\right) \leq e^{\varepsilon} \cdot \operatorname{Pr}\left(G\left(D_{2}\right)=O\right).
\end{equation}
Existing studies have proved that the Laplace and exponential mechanisms are feasible randomized algorithms. For a function $f$ with numerical output, the \rc{\textbf{Laplace mechanism}~\cite{dwork2006calibrating} constructs a corresponding $G_f$ by adding noise sampled from a Laplace distribution $Lap(\frac{\Delta f}{\epsilon})$,
where $\Delta f$ denotes the $l_1$-sensitivity of $f$, that is, the maximum difference between the output of two neighboring datasets.} \rc{The  \textbf{exponential mechanism}~\cite{mcsherry2007mechanism} applies to function $f$ with categorical output, which allows privately selecting the ``best" element from a set.
Assume that $q(D)$ is a designed score function that gives each discrete value a probability; the algorithm $G_f$ provides $\epsilon$-DP if it approximately maximizes the score by returning values from the discrete domain with the probability proportional to $exp(\frac{\epsilon q(D)}{2\Delta q})$, where $\Delta q$ is the $l_1$-sensitivity of $q$.
}



\subsection{\rc{Utility Metrics of Private Visualization}}

\rc{
The dilemma faced by data custodians in~\autoref{sec:exa} motivated us to analyze the utility of privacy-preserving visualizations on data pattern maintenance, which is important for pattern-centric data analysis tasks~\cite{saket2018task}, such as identifying anomalies. Seeking automatic utility evaluation, Bertini et al.~\cite{bertini2006visual} and Behrisch et al.~\cite{behrisch2018quality} studied visual quality metrics and categorized them into} \rc{image-related metrics}, perception-related metrics, and data-related metrics.

Image and perception-related metrics mainly aim to evaluate the effectiveness of a visual representation.
\rc{Treating visualizations as images, image-related metrics inspect the effectiveness through techniques from computer vision and digital image processing~\cite{mojsilovic2001capturing, mojsilovic2004semantic}. Related metrics can detect distribution patterns~\cite{shao2014guided} and evaluate visual clutters~\cite{dasgupta2010pargnostics, ellis2006plot}.} 
Perception-related metrics simulate the results of perception experiments~\cite{cleveland1984graphical}. Those metrics can measure visualizations from the perspectives of memorability~\cite{borkin2015beyond}, aesthetics~\cite{harrison2015infographic}, and engagement~\cite{saket2016beyond}. 
Data-related metrics focus on the statistical characteristics of the data visualized in charts. 
Johansson et al.~\cite{johansson2009interactive} propose user-defined quality metrics to reorder parallel coordinate plots. Scagnostics~\cite{wilkinson2005graph} describes various measures of interest for pairs of variables based on their appearance on a scatterplot. 
Since the proposed approach manipulates the data space with DP techniques, we evaluate the data pattern retention with data-related metrics.


\section{Related Work}\label{sec:relatedWork}

In this section, we summarize related work in privacy-preserving data visualization and interactive \rw{interfaces for} privacy protection.

\subsection{Privacy-Preserving Data Visualization}

Privacy-preserving data visualization~\cite{bhattacharjee2020privacy} aims to make visualizations publicly available without privacy violation. Two approaches are widely used: screen-space and data-space sanitization~\cite{dasgupta2011adaptive}.
Screen-space sanitization leverages visual uncertainty such as a blur or overlap to protect privacy~\cite{chou2019privacy}.
Novel visual designs~\cite{dasgupta2011adaptive,andrienko2016scalable} and evaluation metrics~\cite{dasgupta2013measuring,chou2018empirical} have been proposed to prevent privacy breaches. 
However, there is currently no rigorous privacy definition available for screen-space sanitization~\cite{zhang2020investigating}; thus, its usage is still disputable without extensive evaluation.


Data-space sanitization mainly uses privacy protection algorithms in the data-space and visualizes the resulting data. 
\rc{For example, Lin et al.~\cite{lin2020taxthemis} used $k$-anonymity and numeric variance algorithms to preserve privacy in taxpayers' profiles and transaction records.
Chou et al.~\cite{chou2016privacy, chou2019privacy} applied syntactic anonymity techniques on event sequences and then extended it to social networks. They reduced data-level privacy leaks by modifying graph data such that the nodes and edges satisfy $k$-anonymity and $l$-diversity~\cite{chou2017privacy}.
Oksanen et al.~\cite{oksanen2015methods} introduced the privacy-preserving heat map from mobile sports tracking application data to protect the location privacy of individuals.}

\rc{A more robust and popular approach in data-space sanitization is to use DP because of its resistance to several privacy attacks, such as re-identification~\cite{henriksen2016re}, reconstruction attacks~\cite{dinur2003revealing}, and differencing attacks~\cite{dwork2014algorithmic}. 
In order to protect individual privacy in geolocation data visualization, Ren et al.~\cite{hongde2014differential} proposed IDP-kmeans, which introduces DP in the k-means algorithm to balance privacy disclosure risks and clustering usability.
Zhang et al.~\cite{zhang2016challenges} proposed a privacy-preserving heat map by discretizing the raw data into grid cells and using the Laplace mechanism to add noise to the data point counts in each cell.
They later generalized a DP pipeline for generating privacy-preserving visualizations~\cite{zhang2020investigating}. 
The pipeline includes a DP data publishing algorithm to add calibrated noise to sensitive data and render the privacy-preserving visualizations based on the privatized data.
They proved that if the privatized data satisfies $\epsilon$-DP, the generated visualization also satisfies $\epsilon$-DP according to the DP's post-processing property~\cite{dwork2014algorithmic}.
} 


The DP data publishing methods can be further divided into interactive and non-interactive~\cite{zhu2017differential}. The interactive methods release query answers, such as mean and quantile~\cite{dwork2010differential, wasserman2010statistical}, one by one on-demand. \rc{However, data custodians usually publish multiple charts when they have multidimensional data}. \rc{The interactive methods become inefficient and are prone to over-noising if a variable appears in many charts.} On the other hand, non-interactive approaches produce ``sanitized'' datasets to support subsequent operations~\cite{qardaji2014priview, xiao2010differential}. Data is generated by approximating the marginal distributions with probabilistic graphical models, such as the Bayesian network~\cite{li2014dpsynthesizer} and Markov random field~\cite{mckenna2019graphical}. Deep learning models, such as GAN and auto-encoder, have also been applied, \rw{but their current performance in capturing the correlations in tabular data is not satisfactory~\cite{tao2021benchmarking}}.

\rc{The proposed \textit{PriVis} model belongs to the data-space sanitization approach, which provides a theoretical privacy protection guarantee by DP. We extend a Bayesian network-based DP technique, \textit{PrivBayes}~\cite{zhang2017privbayes}, to generate privatized data in one go. This non-interactive method generates multiple privacy-preserving charts efficiently. Unlike previous works, we also model user preferences for particular patterns as pattern constraints to guide the structure of Bayesian networks, which improves the utility of private visualizations.
}



\subsection{Interactive \rw{Interfaces for} Privacy Protection}

Applying data privacy protection requires a specific understanding of the underlying protection schemes. Interactive techniques are thus developed to make algorithms transparent and understandable. Most of them also provide an interface to perceive better the trade-offs between privacy and utility~\cite{bhattacharjee2020privacy}. Pioneer works generally use conventional anonymization methods. For example, Wang et al.~\cite{wang2017utility} supported users in applying syntactic anonymity to tabular data and interactively evaluating each attribute's distribution loss. They introduced GraphProtector~\cite{wang2018graphprotector} for graph data, which let the data custodian set priorities and inclusion rules for different nodes. Privacy-preserving operations modify these nodes to maintain important structural features such as node degrees, hub fingerprints, and sub-graphs.

\rc{For DP-based interactive systems, a key feature is to interactively optimize the allocation of privacy budgets, usually by displaying the difference between non-private and private results.} For example, PSI($\Psi$)~\cite{gaboardi2016psi} 
shows the absolute and relative error of queries, such as mean and quantile, \rw{for the users' references to assign privacy budgets.}
Overlook~\cite{thaker2020overlook} supports interactive parameter configurations and shows count queries with privacy guarantees. 
ViP~\cite{nanayakkara2022visualizing} visualizes relationships between privacy budgets, accuracy, and disclosure risks with uncertainty visualization. 
These user-centric approaches provide more flexible and granular options for privacy protection, freeing users from tedious data inspection work and improving productivity. 
\rc{However, they need users to set budgets for each query. Users are responsible for carefully maintaining the global allocation scheme by analyzing and predicting the errors in individual query results; otherwise, the privacy budget would easily be exhausted. Our \textit{DPVisCreator} uses a ``sanitized" dataset with DP, allowing users to generate as many privacy-preserving visualizations as they need. \textit{DPVisCreator} also supports users in expressing pattern preferences and visually comparing errors directly in the charts.}

\section{Requirements Analysis}\label{sec:req}


This section summarizes the design requirements obtained through analyzing the example scenario and interviews with privacy practitioners and visualization researchers. \rc{Similar to the trade-offs made by data custodians, we formed the requirements into two perspectives: privacy protection and visualization utility.}


From the \textbf{privacy protection} perspective, when data custodians generate visualizations from raw data, existing practice seldom considers privacy risks. Given the high expressiveness of visualizations, publishing them could lead to unwanted disclosure of sensitive information, similar to publishing the raw data itself.




\begin{enumerate}[label=\textbf{R{\arabic*}}, nolistsep]

\item \rc{\textbf{Recommend privacy protection schemes for publishing visualizations.} Data custodians are aware of the importance of protecting privacy in data visualizations, as strict regulations bond them to keep valuable data safe. Unprotected charts are vulnerable to malicious attacks that could cause severe consequences. However, data custodians generally lack the skills and expertise to derive a persuasive privacy protection scheme, especially for visualizations. In addition, manual configurations from scratch are inefficient for periodic publishing demands. Automated recommendations for privacy protection schemes can significantly improve data custodians' workflow.}

\item \rc{\textbf{Explain the privacy-preserving visualization generation process.} Although many existing tools have provided theoretical guarantees to the protection schemes, privacy practitioners want more evidence to support the results and enhance their trust in the models. Explaining the privacy protection process helps data custodians understand how and to what extent privacy is protected. Therefore, the visualization generation process should be transparent to users and explained intuitively.}

\end{enumerate}

From the \textbf{visualization utility} perspective, data consumers expect to read data patterns and gain insights from the published visualizations. If privacy protection schemes severely distort such data patterns, they are detrimental to the purpose.

\begin{enumerate}[label=\textbf{R{\arabic*}}, nolistsep]
\setcounter{enumi}{2}

\item \rc{\textbf{Explore data patterns of interest.} Multidimensional tabular data contains rich information under different attributes. Data custodians need to explore the publishing dataset from different aspects to identify sensitive attributes and representative data patterns. They evaluate the intrinsic values of data patterns and seek ways to sustain the important ones in the privacy protection process. We should support visual exploration because the final output is in a visual form. 
}


\item \textbf{\rc{Examine different configurations to balance the privacy-utility trade-off.}} \rc{While the automatic generation process can yield privacy-preserving visualizations, the data pattern preference elicited by data custodians might not be sustained. Data custodians should review the protection results to check whether the current privacy budget constraints and pattern preferences produce satisfactory results. Then, they should adjust their preferences and optimize configurations to obtain acceptable results.}



\item \textbf{\rc{Compare different privacy protection schemes.}} \rc{For different analysis purposes, data custodians may indicate different data pattern preferences. They want to compare different privacy protection schemes for an optimal solution to their goal. We should provide quantitative metrics to help them evaluate these schemes. Moreover, multiple charts might be produced under a single privacy protection scheme, and the desired data patterns might not be easily perceived among these charts. Therefore, we should also provide support for qualitative assessments of patterns.}



\end{enumerate}

\section{Our Approach}\label{sec:approach}

In this section, we describe the workflow of our approach and details of the proposed model called \textit{PriVis}. 

\subsection{Approach Overview}
\begin{figure*}[htb!]
 \centering 
 \includegraphics{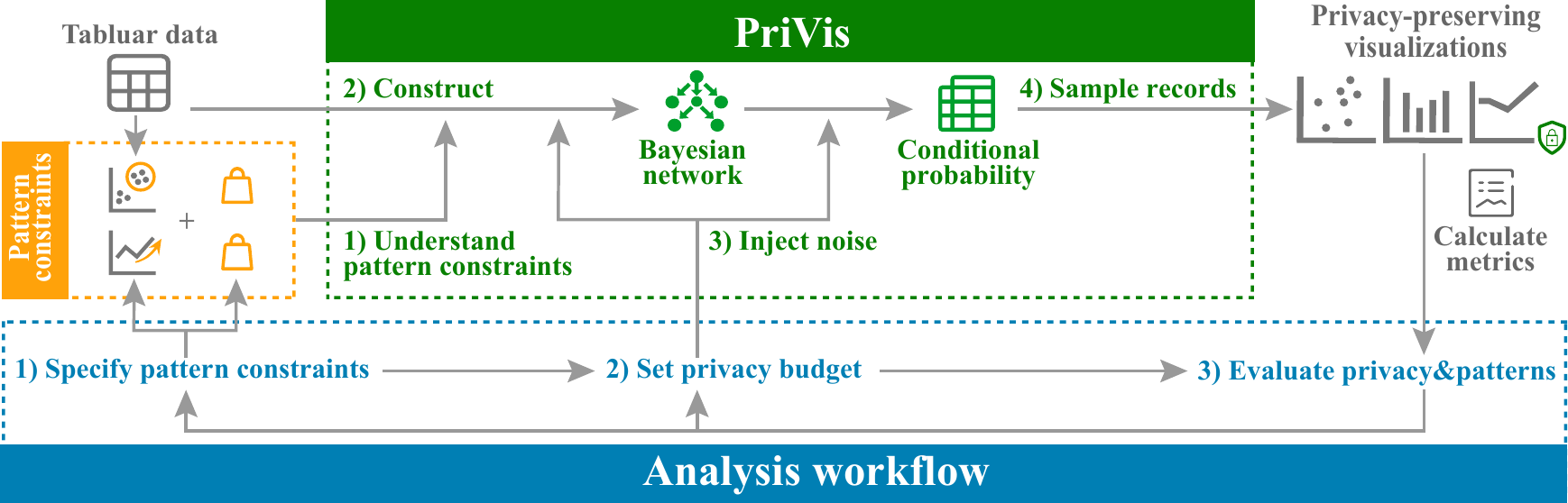}
 \caption{The overview of our approach. Users can select patterns of interest from visualizations of tabular data and specify importance weights. The \textit{PriVis} model considers these pattern constraints and privacy budgets to generate privacy-preserving schemes. The user can evaluate the results based on the visualization charts and utility metrics and return for iterative adjustments.}
 \label{fig:workflow}
\end{figure*}

Our approach lets the user guide the automatic privacy preservation from the utility and privacy perspectives. As shown in \autoref{fig:workflow}, the user first elicits their requirements on the utility by creating a series of pattern constraints. Each \textbf{pattern constraint} consists of a group of data records and a weight of significance. After browsing the data overview, the user creates pattern constraints in two steps: 1) select a data group by marking the corresponding visual elements from a chart, and 2) set the corresponding weight. Based on these pattern constraints, \textit{PriVis} recommends private visualizations to the user, which would then be evaluated to balance the trade-off between privacy and utility. If none of the recommendations are satisfactory, the user should specify other pattern constraints or set another privacy budget. 

\subsection{Our model}
As mentioned in Sec.~\ref{sec:exa}, significant patterns need to be preserved in private visualizations. Although existing approaches, like \textit{PrivBayes}~\cite{zhang2017privbayes}, can provide privacy preservation for tabular data, none of them consider patterns in visualizations. To address this issue, we propose \textit{PriVis}, which extends \textit{PrivBayes} to support pattern constraints. In this section, we first introduce \textit{PrivBayes} and then describe \textit{PriVis}. For clarity of description, we have listed the notations in \autoref{tab:notations}.

\subsubsection{PrivBayes}
\textit{PrivBayes} is a DP algorithm for publishing high-dimensional datasets. Due to the curse of dimensionality~\cite{koppen2000curse}, processing high-dimensional data tables often creates uncontrollable noise, resulting in the degradation of usability. To solve this problem, \textit{PrivBayes} first abstracts the high-dimensional dataset into low-dimensional marginal distributions, abbreviated as \textit{marginals}. Sampling noise from the marginals, \textit{PrivBayes} generates privacy-preserved high-dimensional data.

To calculate marginals, \textit{PrivBayes} employs the Bayesian network, a probabilistic graphical model that summarizes tabular data via a directed acyclic graph (DAG). In a Bayesian network $N$, each node represents an attribute, and each directed edge describes the conditional probability between two attributes. A linear ordering of vertices ($X_1, X_2,\dots, X_d$) can be obtained by topological sorting.
Formally, $N$ approximates the probability distribution of high-dimensional data by a series of distributions of attribute-parent (AP) pairs $(X_i, \Pi_i)$, having
\begin{equation}
\begin{aligned}
\operatorname{Pr}_{N}(A) &=\operatorname{Pr}\left(X_{1}, X_{2}, \ldots, X_{d}\right)
=\prod_{i=1}^{d} \operatorname{Pr}\left(X_{i} \mid \Pi_{i}\right),
\end{aligned}
\end{equation}
where $X_i$ represents the $i$-th attribute in $A$ and $\Pi_i$ is the set of parents of node $X_i$ in network $N$.

The approximation degree of $N$ to the original data distribution $Pr[A]$ can be measured by the \textit{KL divergence}, which is defined as
\begin{equation}
D_{K L}\left(\operatorname{Pr}(A) \| \operatorname{Pr}_{N}(A)\right)=-\sum_{i=1}^{d} I\left(X_{i}, \Pi_{i}\right)+\sum_{i=1}^{d} H\left(X_{i}\right)-H(A),
\end{equation}
where $H(X)$ represents the entropy of the attribute $X$ over its domain, and $I(\cdot, \cdot)$ represents the mutual information between the two variables. When the dataset $D$ is given, $\sum_{i=1}^{d} H\left(X_{i}\right)-H(A)$ is also fixed. The construction of the Bayesian network is transformed into an optimization problem that maximizes the mutual information between all AP pairs. 

To protect privacy, \textit{PrivBayes} introduces the exponential mechanism to select AP pairs by using mutual information as a score function and generates marginals with Laplace noise. The released dataset can then be synthesized by sampling from the marginals. According to the post-processing property~\cite{dwork2014algorithmic} of DP, \textit{PrivBayes} has a privacy guarantee.



\begin{table}[htb]
  \centering
  \caption{Notation definitions used in this paper.}
  \label{tab:notations}
  \begin{tabular}{p{0.1\linewidth} p{0.8\linewidth}   }
  \toprule
  \textbf{Notation}  & \textbf{Description}  \\ 
  \midrule
      $D$ & A multidimensional table including sensitive information \\
      $n$ & The number of records in $D$   \\
      $A$ & The set of attributes in $D$  \\
      $d$ & The number of attributes in $A$ \\
      $N$ & A Bayesian network over $A$ \\
      $X$ & The nodes in Baysian network \\
      $\Pi$ & The set of parents of nodes in Bayesian network\\
      $Pr(A)$ & The distribution of records in $D$ \\
      $Pr_N(A)$ & An approximation of $Pr(A)$ defined by $N$ \\
      $dom(X)$ & The domain of attribute $X$ \\
      $\mathbbm{1}_{P_k}$ & Indicator function for the $k$-th pattern \\
      $w(k)$ & The importance weight for the $k$-th pattern \\
      $MW(r)$ & The mixture weight value of a record $r$ \\
  \bottomrule
  \end{tabular}
\end{table}

\subsubsection{PriVis}

\textit{PriVis} generates privacy-preserving visualizations in four steps.

\textbf{Step 1: Understand pattern constraints.} 
Users may select multiple patterns from different charts, represented as $P = \{P_1, P_2, \cdots, P_s\}$. Each pattern $P_i$ corresponds to a subset of data records in the whole data $D$. For example, as mentioned in Sec.~\ref{sec:exa}, the data custodian needs data records with both age and illness attributes from the hospital data to generate a bar chart. To emphasize the illness risk in people over 60, the data custodian can specify the pattern by selecting bars corresponding to patients over 60 and setting a weight. 
User preferences for different patterns can be recorded as different subsets of data records with corresponding weight $W = \{w_1, w_2, \cdots, w_s\}$. 
For a data record $r$, its final weight needs to consider its occurrence under each pattern. To this end, we give each record an initial weight of 1 and then add additional weights of $w_k$ for each record $r$ in pattern $P_k$. The mixture weight $MW(r)$ of data record $r$ is defined as:

\begin{equation}
MW(r) = \sum_{k=1}^{s} w_k \mathbbm{1}_{P_k}(r) + 1,
\end{equation}


where $\mathbbm{1}_{P_k}$ is indicator function which maps record in pattern $P_k$ to 1, and all others to 0. Therefore, we translate the different pattern constraints into a mixture weight assigned to each record.

\textbf{Step 2: Construct a Bayesian network with pattern constraints.} To preserve user-specified patterns, \textit{PriVis} emphasizes the correlations between the corresponding records when constructing the Bayesian network. Specifically, when selecting AP pairs, \textit{PriVis} not only employs the exponential mechanism but also replaces the mutual information in the PrivBayes network with the weighted mutual information: 

\begin{equation}
\begin{split}
I_w(X, \Pi, MW)= 
\sum_{x \in \operatorname{dom}(X)} \sum_{\pi \in \operatorname{dom}(\Pi)} \frac{\sum\limits_{r \in x \cap \pi}MW(r)}{|x \cap \pi|} \\ \operatorname{Pr}(X=x, \Pi=\pi)
\log \frac{\operatorname{Pr}(X=x, \Pi=\pi)}{\operatorname{Pr}(X=x) \operatorname{Pr}(\Pi=\pi)}.
\end{split}
\end{equation}

With mixture weights, the constructed Bayesian network can better preserve dependence among user-selected data. Therefore, the patterns consisting of these records can be better maintained in visualizations.


\textbf{Step 3: Inject noise to the Bayesian network.} The constructed Bayesian network provides conditional probability distributions $Pr(X_i|\Pi_i)(i\in[1,d])$ to approximate the distribution of the high-dimensional dataset. To inject noise, \textit{PriVis} first calculates noisy distribution $Pr^{*}(X_i, \Pi_i)$ for any $i\in[k+1, d]$ by adding Laplace noise to joint probability distribution $Pr(X_i, \Pi_i)$. Then, the conditional distribution $Pr^{*}(X_i|\Pi_i)$ can be derived. The $Pr^{*}(X_i|\Pi_i)(i\in[1,k])$ can thus be directly obtained from $Pr^{*}(X_{k+1}|\Pi_{k+1})$.


\textbf{Step 4: Sample records to generate private datasets.} 
According to the topological order of the Bayesian network, $X_i(i\in[1,d])$ can be sampled in increasing order of $i$ based on the noisy conditional distribution. For example, when all parent attributes of $X_j(j\in[2,d])$ are sampled, $X_j$ can be sampled by conditional probability $Pr^{*}(X_j|\Pi_j)$. 
The sampled data can then be used for publishing visualizations.

\textbf{Privacy Analysis.} In our approach, the data custodian is located in a trust zone to elicit pattern preferences before the model performs privacy learning. The model is guided towards the right direction without additional privacy budgets. In the \textit{PriVis} model, access to the original data is required for steps 2 and 3. In step 2, the model uses the exponential mechanism to privately select AP pairs, which converts selection from multiple candidate pairs into probabilistic sampling, which consumes an $\epsilon_1$ privacy budget. In step 3, Laplace noise is added to the conditional probability distributions, which consumes an $\epsilon_2$ privacy budget. According to the composition property~\cite{dwork2014algorithmic} of DP, the resulting visualizations satisfy the $\epsilon$-DP, where $\epsilon = \epsilon_1 + \epsilon_2$.

\section{System Design}\label{sec:design}

In this section, we present a system overview and then introduce the details of the visual design and interaction of \textit{DPVisCreator}~\footnote{\url{https://github.com/wanghuanliang/DPVisCreator}}

\begin{figure*}[htb!]
 \centering 
 \includegraphics[width=2\columnwidth]{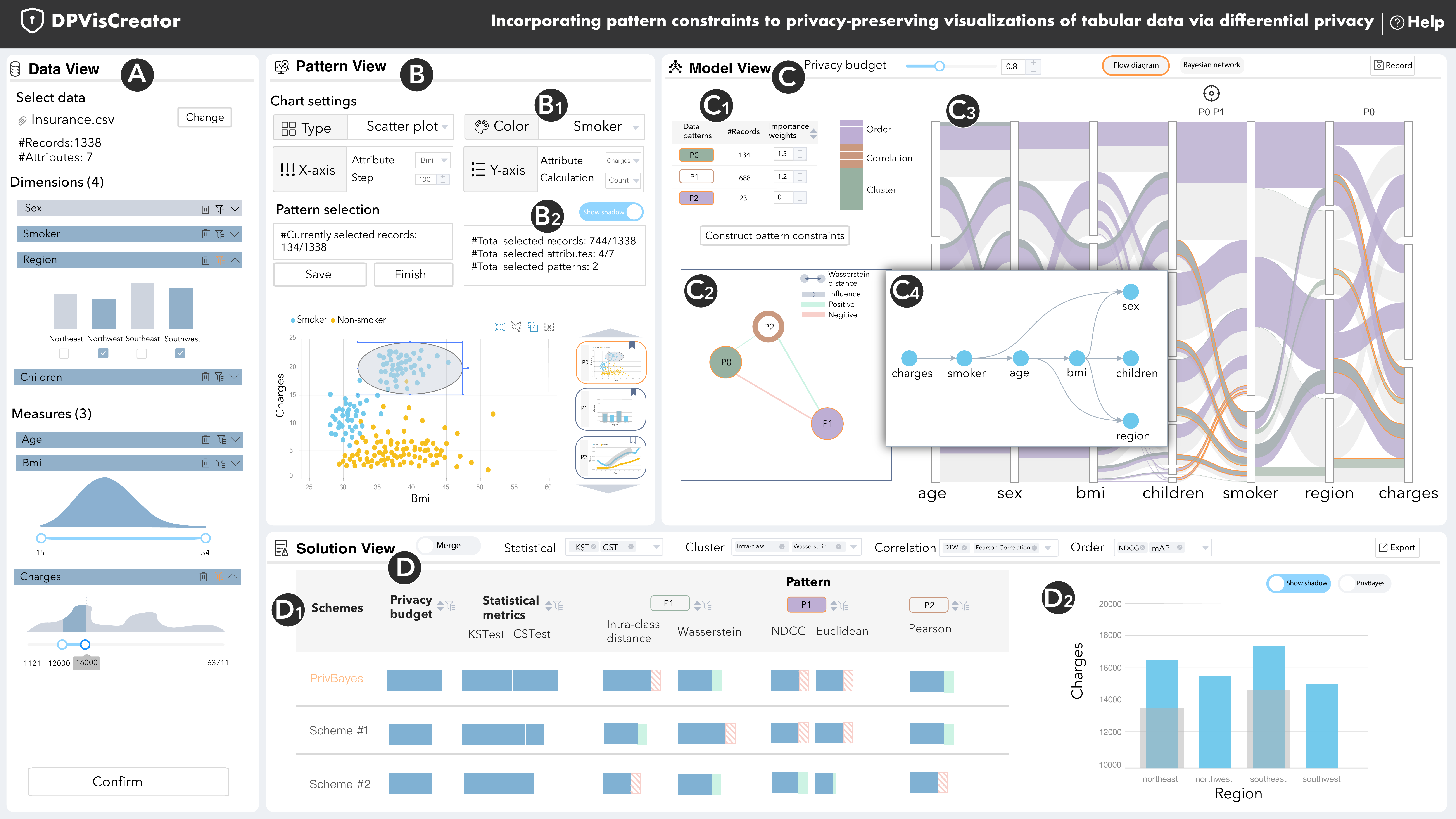}
 \caption{\textit{DPVisCreator} facilitates publishing privacy-preserving visualizations. (A) The \textit{Data View} displays the distribution of attributes. (B) The \textit{Pattern View} supports the pattern specification. (C) The \textit{Model View} provides the relationship between data constraints and the underlying Bayesian network structure. (D) The \textit{Solution View} assists comparison and assessment of privacy-preserving schemes.}
 \label{fig:sys}
\end{figure*}

\subsection{System Overview}

To meet the requirements in Sec.~\ref{sec:req}, we designed \textit{DPVisCreator} with four main views, as shown in~\autoref{fig:sys}: the \textit{Data View}, the \textit{Pattern View}, the \textit{Model View}, and the \textit{Solution View}. 
We describe an analysis flow to illustrate how the four views help data custodians generate privacy-preserving visualizations with important data patterns. 
\rc{
A data custodian first loads a dataset and looks for sensitive attributes in the \textit{Data View}~(\autoref{fig:sys}-A), which shows the general data characteristics. He selects attributes of interest and specifies the desired chart settings in the \textit{Pattern View}~(\autoref{fig:sys}-B). He can also highlight the data patterns of interest with the data selection tools~(\textbf{R3}). Then, these preferences are processed by \textit{PriVis} to propose a privacy protection scheme~(\textbf{R1}). 
}

\rc{
The recommended scheme is visualized in the \textit{Model View}~(\autoref{fig:sys}-C), with its structure and impact explained~(\textbf{R2}). He can examine the relationships between different pattern constraints and adjust the importance weight for different results~(\textbf{R4}). 
He turns to the \textit{Solution View}~(\autoref{fig:sys}-D) to evaluate the privacy protection scheme in terms of privacy and utility~(\textbf{R4}). All generated schemes are recorded in this view for comparison~(\textbf{R5}). He can revisit the \textit{Model View} to generate more candidates and strike a desirable balance between privacy and utility~(\textbf{R4}). Iterative optimizations can find a privacy-preserving scheme that maintains important patterns with an acceptable privacy budget.}

\subsection{Data View}

The \textit{Data View}~(\autoref{fig:sys}-A) provides an overview of the loaded dataset by showing its attributes and the corresponding data distribution. It supports data range filtering and allows users to choose attributes of interest for subsequent analysis (\textbf{R3}).

The view allows the user to upload tabular data with the number of records and attributes displayed below.
\rw{
Each attribute can expand to show its data distribution and be selected accordingly.
Selected attributes are colored in blue, and unselected ones in gray.
For categorical attributes, their data records are aggregated and displayed in bar charts. 
Every unique data value can be selected by checking the selection box underneath. 
For numerical attributes, their probability density distributions are first calculated using kernel density estimation and then presented using a curve plot. 
A slider indicates the data value range and is used to filter the range of interest. 
After finishing the data selection process, users can click on the confirm button to proceed.
}

\subsection{Pattern View}

The \textit{Pattern View}~(\autoref{fig:sys}-B) consists of two parts: (1) The \textit{chart settings panel} lets users configure visualization charts; (2) The \textit{pattern selection view} supports the interactive selection of data patterns~(\textbf{R3}).

The \textit{chart settings panel}~(\autoref{fig:sys}-B1) lets users select the chart type, color, and attributes for the x- and y-axis of the charts. Three common charts (i.e., scatter, line, and bar charts) are supported.
Users can quickly specify chart types by mapping selected attributes to appropriate visual encodings via drop-down boxes.
In order to reduce the visual clutter for large amounts of data, the view supports step adjustment for the x-axis and aggregation calculation for the y-axis. 

The \textit{pattern selection view}~(\autoref{fig:sys}-B2) contains an information card, a visualization chart, and a pattern list.
\rw{
Users can select the data patterns of interest from the visualization chart, which is rendered according to the above configuration. 
Due to the different shapes of the visualized data patterns, we provide a set of interactive selection methods.
For cluster patterns in scatter plots, use the box and lasso to select regions of interest.
For correlation patterns in line charts, horizontal selection indicates intervals that match a specific trend.
For order patterns in bar charts, the bar being clicked is chosen.
Gray backgrounds are overlaid on all selected areas.
The information card shows the statistics of current and all patterns, and the pattern list records all selected patterns to track analytic provenance. 
Users can add the current pattern to the list and revisit previously saved patterns by clicking it.
}

\subsection{Model View}

The \textit{Model View} contextualizes the privacy-preserving process to foster users' understanding and supports users in making trade-offs between data pattern maintenance and privacy protection.
The \textit{Model View}~(\autoref{fig:sys}-C) contains four parts: (1) The \textit{weight configuration panel} lets users adjust importance weights to each pattern~\textbf{(R4)}; (2) The \textit{pattern constraint relationship} represents the overall impact between different pattern constraints~\textbf{(R4)}; (3) The \textit{pattern constraint flow} shows the detailed pattern constraint distribution on each attribute and highlight them for comparisons \textbf{(R4)}; (4) The \textit{Bayesian network view} visualizes the structure in the underlying \textit{PriVis} model \textbf{(R2)}. The latter two views can be switched with a button. 

\rw{
The \textit{weight configuration panel}~(\autoref{fig:sys}-C1) uses a stacked bar chart to provide an overview of the importance weight distribution. 
Colors encode the pattern constraint's type, having green for the cluster, brown for correlation, and purple for order. 
This color encoding is applied throughout the system. 
The corresponding pattern ID involving data amount and importance weights are displayed on the data pattern list and hovers on the corresponding bar. 
Users can adjust the corresponding pattern constraint weights in the data pattern list. 
}

\rw{
The \textit{pattern constraint relationship}~(\autoref{fig:sys}-C2) encodes pattern constraints as circles. 
The position of a circle is determined by multidimensional scaling (MDS)~\cite{cox2008multidimensional}, where the Wasserstein distance~\cite{vallender1974calculation} is used to measure the similarity between patterns. 
The projection reflects the differences in distributions. 
Edges between circles encode the impact of different pattern constraints by affecting the network structure.
Their colors encode the type of influence of changing weights. 
Green edges imply positive correlations, where increasing one pattern's weight will promote the other.
In contrast, red represents negative correlations that the other pattern will be weakened. 
The magnitude of influence is encoded by the thickness of the edge, which is calculated by the sum of the difference between the intersection and the symmetric difference of AP pairs. 
Clicking a pattern constraint will highlight it and its adjacent edges, together with the corresponding entities in other views.}

The \textit{pattern constraint flow}~(\autoref{fig:sys}-C3) improves the basic Sankey diagram with sub-flows representing constraints in each attribute. The x-axis lists all attributes, and the y-axis indicates the discrete interval of each attribute. \rc{Continuous attributes are discretized by $k$-means with the elbow method~\cite{marutho2018determination}, while categorical attributes use their discrete domain as the interval.} The length along the y-axis represents the amount of data belonging to this interval. \rc{The specific interval information can be viewed by hovering.} Different flows represent different data distributions, and the width of a flow is proportional to its data volume. By default, all data flows are grayed out to provide a clear background, while the selected data pattern will be highlighted correspondingly for a more intuitive comparison.

The data flow between attributes follows three visual patterns: focus, convergence, and divergence. Focus means that data flowing from a single interval mostly goes to a specific interval of another attribute; convergence means that data flowing from multiple intervals goes to the same interval; divergence means data flowing from a single interval goes to multiple intervals. This difference in data distribution reflects, to some extent, the impact of different data patterns. Suppose that two data patterns have the same attribute pairs of the focus relationship. In this case, they have a high probability of choosing the same AP pairs when constructing the Bayesian network; that is, the network can be a good approximation of the data distribution of both of them. If two data patterns have a significant difference in the attribute pairs of the focus relationship, increasing the weight of one pattern will likely reduce the Bayesian network's approximation effect on the other pattern.

The \textit{Bayesian network view}~(\autoref{fig:sys}-C4) shows the structure of the \textit{PriVis} model. Since the network is a DAG, we use a hierarchical layout\rc{~\cite{battista1998graph}} to show the dependencies of attributes. Specifically, the depth of each node is obtained by topological sorting. We draw each layer in turn along the x-axis, and nodes in the same layer are evenly distributed along the y-axis. Nodes represent attributes, and edges reflect conditional dependencies.
The probability density distributions before and after noise addition are displayed by superposition when hovering over a node, making it easy to compare and understand the exact degree of privacy protection.

\subsection{Solution View}

The \textit{Solution View}~(\autoref{fig:sys}-D) measures privacy-preserving schemes from different perspectives~\textbf{(R4)} and shows comparisons for selecting an appropriate scheme~\textbf{(R5)}. It contains two parts: (1) The \textit{schemes ranking list} displays all privacy-preserving schemes. Inspired by lineup~\cite{gratzl2013lineup}, the schemes' metrics are organized into a multidimensional table. (2) The \textit{pattern comparison view} uses the superposition technique of visual comparisons~\cite{gleicher2017considerations} to highlight the scheme's effect.

The \textit{schemes ranking list}~(\autoref{fig:sys}-D1) shows evaluation metrics for each privacy protection scheme in detail, including privacy budgets, statistical indicators, and pattern retention metrics. 
KSTest~\cite{berger2014kolmogorov} and CSTest~\cite{greenwood1996guide} analyze the differences in statistical properties of data.
As to utility, suitable metrics are adopted for different patterns, such as Wasserstein distance for the cluster, dynamic time wrapping (DTW) for correlation, and Euclidean distance for order.
These metrics reflect the degree of pattern retention. Users can choose different metrics to measure the scheme's utility according to their needs. 
The length of the horizontal bar encodes the metrics' value.
We explicitly encode the differences in the metrics before and after privacy preservation for comparing pattern-related metrics. 
The increased proportion is represented by green bars, while the decreased proportion is represented by red striped bars. 
Users can rank schemes based on metrics with sorting and filtering functions.

The \textit{pattern comparison view}~(\autoref{fig:sys}-D2) shows the visualization chart after privacy protection. 
The gray border represents the previous selection area, and the gray visual elements represent the original data. This visual comparison helps users qualitatively perceive the difference in visualization utility before and after. Users can also switch off the pattern constraints, i.e., to compare the chart with the baseline.

\section{Evaluation}\label{sec:evaluation}
We implemented a quantitative experiment to assess \textit{PriVis} and described two case studies to verify the effectiveness of \textit{DPVisCreator}. We finally report subjective feedback gathered from \rc{domain experts}.

\subsection{Quantitative Evaluation}

\textbf{\rc{Experiment settings.}} \rc{The purpose of this experiment is to test the ability of \textit{PriVis} to maintain preferred data patterns in visualization while preserving privacy. We chose \textit{PrivBayes}~\cite{zhang2017privbayes} as the baseline method, which has been shown to perform well on tasks such as statistical queries and classification~\cite{tao2021benchmarking}. We compare \textit{PriVis} and \textit{PrivBayes} on three types of data patterns separately under various privacy budgets and also evaluate the effect of \textit{PriVis} on multiple patterns under different weight configurations. The same condition has 25 replicate runs to mitigate the effects of randomness introduced by noise. The detailed experimental information can be found in the \url{https://osf.io/ugw29/}.
}

\rc{\textbf{Datasets} We tested these two methods on five datasets with different distribution characteristics, numbers of data records, and attribute types.
\textit{Shopping}~\footnote{\url{https://archive.ics.uci.edu/ml/datasets/Online+Shoppers+Purchasing+Intention+Dataset}}, \textit{Adult}~\footnote{\url{https://archive.ics.uci.edu/ml/datasets/adult}} and \textit{Bank}~\footnote{\url{https://archive.ics.uci.edu/ml/datasets/bank+marketing}} datasets are from the UCI machine learning repository, and \textit{Insurance}~\footnote{\url{https://www.kaggle.com/datasets/teertha/ushealthinsurancedataset}} and \textit{Loan}~\footnote{\url{https://www.kaggle.com/datasets/burak3ergun/loan-data-set}} from Kaggle.
}

\rc{\textbf{Evaluation metrics}. We chose the corresponding metrics for different data patterns. For cluster, all data before and after privacy protection are used to calculate the metrics since the selection area may be inaccurate and the cluster position could move out of the selected region due to noise. For correlation and order, data in the user-selected range are used. The specific metrics are as follows:}

\begin{itemize}[nolistsep]
  \item \textbf{Cluster.} The Wasserstein distance~\cite{vallender1974calculation} is used to measure the distribution distance of data points in scatter plots.
  \item \textbf{Correlation.} The difference of Pearson correlation coefficients~\cite{benesty2009pearson} is used to measure the change in trend, and DTW~\cite{Muller2007} is applied to measure the distance between the selected lines.
  \item \textbf{Order.} The NDCG~\cite{busa2012apple} is used to reflect the correctness of ranking, while the amount of variation between selected bars is calculated by the Euclidean distance.
\end{itemize}

\begin{figure*}[htb!]
    \centering 
    \includegraphics[width=1.99\columnwidth]{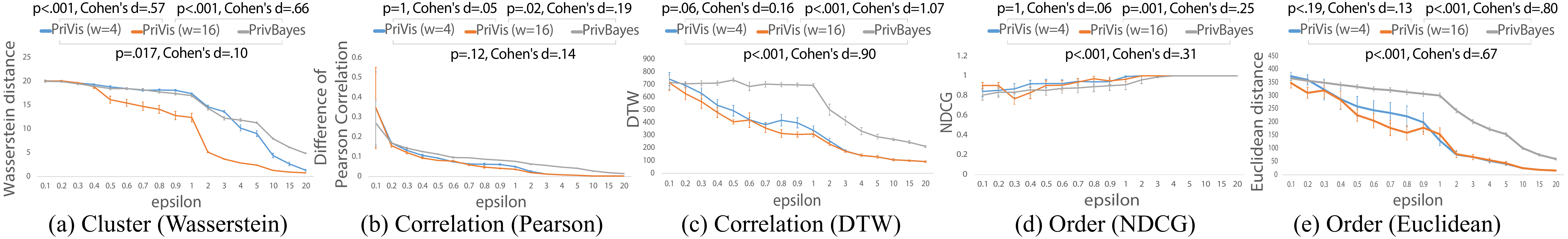}
    \caption{The comparison of \textit{PriVis} (weight=4/16) and PrivBayes~\cite{zhang2017privbayes} in maintaining three data patterns under various epsilon conditions on the adult dataset. Except for NDCG, the smaller the value, the better. The p-values and effect sizes are labeled next to the legend. Error bars represent 95\% confidence intervals.}
    \label{fig:comparison}
\end{figure*}

\textbf{Results.} \rc{We conducted ANOVA and posthoc tests with Bonferroni correction to assess the effectiveness of the different methods for pattern maintenance. As shown in~\autoref{fig:comparison}, for three types of patterns, different methods perform similarly when the degree of privacy protection is strict ($\epsilon < 0.5$), mainly due to a large amount of noise that severely obscures the original data. When the protection level is moderate ($\epsilon \in [0.5,5]$), \textit{PriVis} outperforms \textit{PrivBayes}, and the pattern retention effect can be improved by increasing the weights, which shows the improvement of the visualization utility by optimizing the bayesian network structure. When the privacy need is relaxed ($\epsilon > 5$), noise interference is less, and the difference in the performance of methods gradually narrows. The maintenance effect of cluster is more significant, while correlation and order are mainly reflected in the reduction for distance deviation. As shown in~\autoref{tab:experiment2}, we compare the performance of \textit{PriVis} and \textit{PrivBayes} in generating three charts with different patterns on the adult dataset. The results show that the maintenance of the different data patterns can be improved by adjusting the corresponding weight configuration of \textit{PriVis}.}

\begin{table}[htb]
  \centering
  \caption{The mean of pattern metrics for different schemes. Metrics are reported for each pattern in the order mentioned above. ($\epsilon=2$)}
  \label{tab:experiment2}
  \begin{tabular}{p{0.17\linewidth}{l} p{0.2\linewidth}{r} p{0.15\linewidth}{r} p{0.12\linewidth}{r} }
  \toprule
  \textbf{Schemes} & \textbf{Cluster}  & \textbf{Correlation}  & \textbf{Order} \\ 
  \midrule
      \multirow{2}{0.17\linewidth}{PrivBayes} & \multirow{2}{0.2\linewidth}{264.05} & 1.61 & 0.86 \\
                   &  & 281.20 & 531.40 \\
      \cline{1-4}
      \textit{PriVis} & \multirow{2}{0.2\linewidth}{277.92} & 1.49 & \textbf{0.96}\\
         W=(1, 1, 4)  &  & 279.20 & \textbf{510.80} \\
         \cline{1-4}
      \textit{PriVis} & \multirow{2}{0.2\linewidth}{287.08} & \textbf{1.13} & 0.88\\
        W=(1, 9, 4)  &  & \textbf{252.80} &  514.20 \\
        \cline{1-4}
      \textit{PriVis} & \multirow{2}{0.2\linewidth}{\textbf{256.85}} & 1.27 & 0.88 \\
      W=(9, 4, 1)  &  & 287.40 & 517.00 \\
  \bottomrule
  \end{tabular}
\end{table}

\rc{\textbf{Limitations.} Although these results demonstrate the advantages of \textit{PriVis} in pattern retention, there exist some factors that can affect the validity, such as the data selection and weights settings. Since \textit{PriVis} improves the pattern utility by influencing the network structure, it is difficult to change the direction of structure learning when the amount of selected data is small and the weights are low. Also, when the distribution of the selected data and the whole data are similar, the current network structure may already be optimal and cannot be further improved. The results of \textit{PriVis} and \textit{PrivBayes} are comparable in the above conditions. In particular, since DP introduces random noise, \textit{PriVis} may be less effective when epsilon is small.
}

\subsection{Case Studies}
We introduce two case studies to show the privacy-preserving visualizations produced by \textit{DPVisCreator}.

\subsubsection{Health Insurance Dataset}
As shown in \autoref{fig:sys}, the dataset contains attributes for 1,338 U.S. insured persons, including age, sex, BMI, number of children, smoker, region, and charges. The data custodian wanted to analyze the impact of different attributes on premium charges, so he uploaded the table into the \textit{Data View}. The attribute distribution revealed that the costs mainly concentrate around 15k and 40k. To investigate the reasons for this difference, he analyzed the relationship between charges and other attributes in charts of the \textit{Pattern View}. He quickly generated a scatter plot of BMI and charges by selecting the x- and y-axis attributes via the drop-down boxes. He found that points with BMI between 30 and 45 and charges between 30,000 and 50,000 formed a cluster. After color encoding the attribute of smoker, he found all the people in the cluster were smokers, reflecting those obese people who smoked had high charges. It was an important finding, so he saved it as data pattern P0. He then intended to see if there was a geographical difference in charges by generating a bar chart of regions and charges. He found that the southeast and northeast regions ranked first and second in average charges and were significantly higher than other regions. This data pattern P1 also reflected the differences in medical conditions across regions and was therefore saved.

If original charts were published directly, an attacker could directly read sensitive information from charts or infer the privacy of the insured through background knowledge. Therefore, he decided to treat them with privacy protection. In the \textit{Model View}, he configures the same importance weight of 0.5 for both patterns to show his preference. Then, he clicks the record button to have the system generate a privacy protection scheme\#1 accordingly. By looking at the \textit{Solution View}, he found that the order of the selected bar was well maintained, but the cluster in the scatter plot was not concentrated. So he returned to the \textit{Model View} to make adjustments.
The \textit{pattern constraints relationship} shows a negative influence edge between P0 and P1, and the \textit{pattern constraints flow} indicates that their data flows co-focus on a small percentage of attributes, which prevents the network structure from being able to approximate the distribution of the two data patterns simultaneously. To ensure the effectiveness of the cluster pattern, he adjusts the weights of cluster and order to 1 and 0.1, respectively, thus obtaining a new scheme\#2. As shown in~\autoref{fig:teaser}, the effect of the cluster had improved, and the order had not been affected much, which was a satisfactory result. To further confirm, he looked at the network structure of the \textit{PriVis} model for scheme\#1 and scheme\#2 and found that its structure had changed, which in turn led to the change. The quantitative metrics also reflected the utility of these visualization charts, which he was satisfied with and finally exported.

\begin{figure*}[htb!]
  \centering 
  \includegraphics{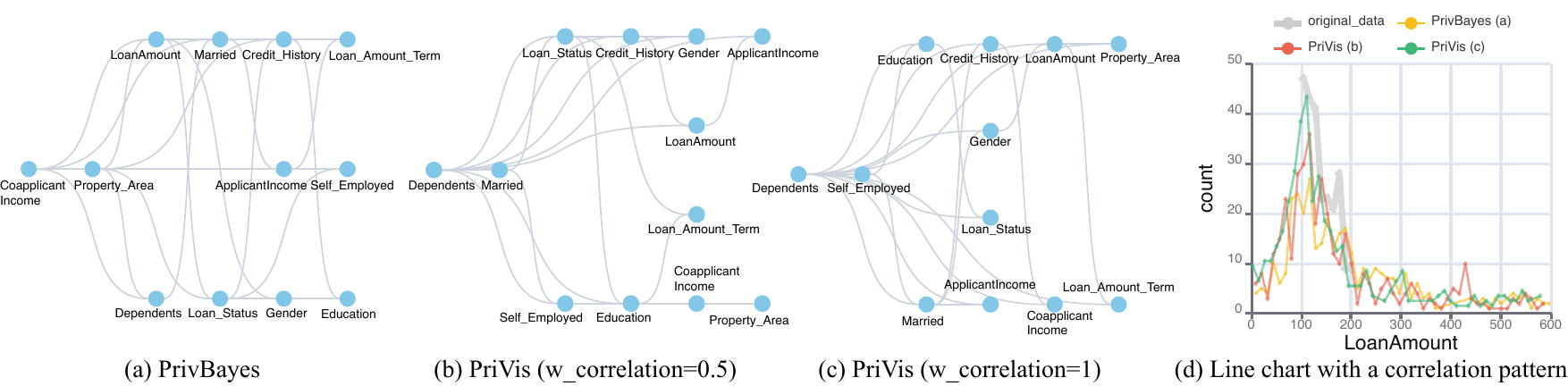}
  \caption{The development of the Bayesian network in Case 2. (a-c) The Bayesian network structures of different schemes. (d) The correlation maintenance is getting better with the corresponding weight increase.}
  \label{fig:case2}
\end{figure*}

\subsubsection{Loan Dataset}

As shown in \autoref{fig:case2}, the dataset contains detailed information about customers applying for loans, including gender, marital status, education, etc. The data custodian wanted to analyze which factors could help identify customer segmentation and produce reports to aid loan qualification approval. In order to know the number of loan origination for different loan amounts, a line chart was generated. He found that the loan amount around 100 have the highest origination number, but the number declined rapidly with the loan amount increased, as shown in \autoref{fig:case2}-(d). He saved the pattern as P0. He further analyzed the income of co-applicants whose loan amount ranged from 100-200. By quickly generating a scatter plot and color encoding education, he found that co-applicants who have not graduated formed a cluster. He recorded this pattern of the stratification of the loan population as P1. Similarly, he reviewed a bar chart showing the number of dependents and loan amount and saved another data pattern P2 which indicates that the fewer relatives, the lower the loan amount. 

Using the three pattern constraints, the data custodian proceeded to privacy protection processing before publishing the visualizations. The \textit{pattern constraints relationship} revealed that P0 had a positive influence on both P1 and P2, so he increased the weight of P0 from 0.5 to 1. The results revealed that the adjustment led to an improvement in the metrics for all three patterns. The underlying Bayesian network (\autoref{fig:case2}-(a-c)) 
showed that the network structure changed as the weights increased, leading to an improvement of correlation in the private line chart. This case demonstrated that \textit{DPVisCreator} could help users adjust and optimize privacy protection schemes.

\subsection{Expert Reviews}

\rc{\textit{DPVisCreator} was reviewed by four domain experts ($E_A$-$E_D$) who were potential target users with basic knowledge in DP and data visualization.}

\textbf{\rc{Procedure.}} \rc{For each interview, we first introduced our project background and requirements (10 minutes). Then the analysis process of the system was demonstrated through case studies (20 minutes), followed by free exploration in a think-aloud manner (15 minutes). Finally, we collected the feedback (15 minutes)} summarized below.

\textbf{\rc{Visualization and interaction.}} \rc{All experts agreed that \textit{DPVisCreator} meets analysis requirements. We have observed that they might not immediately remember all the visual encodings and interactions when first encountering the system. However, after a demo explanation, they were able to understand. $E_A$ appreciated that the multiple selection options (click, box, and lasso) in the \textit{Pattern View} gave him the freedom to express a preference on specific data. At the same time, he admitted that this selection might also have errors and needs to be combined with the filtering function in the \textit{Data View} to select the important pattern more accurately.} For the \textit{Model View}, $E_A$ and $E_C$ commented that \textit{``the pattern flow visually show the `flow' of different patterns across attributes, making it easy to see differences or consistency.''} \rc{$E_B$ adds, ``Stacked bar chart provide a nice overview to help adjust the weights interactively, compensating for the current one-size-fits-all privacy protection approach"}. \rc{Regarding the \textit{Solution View}, there are different comments.} \rc{$E_A$ believed that \textit{``Quantitative metrics-based \textit{schemes ranking list} is consistent with most existing data-centric privacy processes, which is more familiar for me to use.''}} $E_D$ emphasized that \textit{``overlaying data before and after protection with the user-selected area provides a visual reference for comparing pattern retention...\rc{It might be more intuitive to compare charts of different schemes with a small multiple techniques.}''}

\textbf{System usability.} \rc{All experts agreed that the overall analysis flow of the system was clear.} The combination of automated model and interactive visualizations helps them publish privacy-preserving visualizations more efficiently and make a trade-off between data pattern maintenance and privacy protection. $E_B$ mentioned that \textit{``previous approaches to privacy protection required writing a lot of code and manually checking the results, which are time consuming and laborious. \rc{This user-centric approach to privacy was much more controllable, allowing users to compare and adjust privacy protection schemes interactively.}''} \rc{$E_C$ said, \textit{``If I need to visualize some sensitive data for the public, I would be interested in using this system for processing because it makes the complex DP model more transparent and intuitive"}}. \rc{Both $E_A$ and $E_B$ pointed out that adjusting both importance weights and privacy budget parameters required some expertise. If the parameters can be set automatically, it will help to simplify the process. They also admit that in current practice, finding an optimal privacy protection scheme was inherently tricky, and joining the human involvement would help to find an acceptable scheme more efficiently.}

\textbf{Suggestions.} Experts also provided valuable suggestions based on our research. $E_D$ suggested that data tables before and after privacy protection could be displayed to provide more detailed information. $E_A$ and $E_C$, on the other hand, noted that when the system handles more than 15 attributes, it does not give timely interactive feedback. Hence the computational efficiency needs to be improved.
There were also suggestions to include domain-specific evaluation metrics. \textit{``\rc{The definition of utility is task-specific. The current evaluation metrics are the basis for many analysis tasks, but more users could benefit from support for customizable and more complex utility evaluations. For example,} metrics for periodic patterns are important when publishing time-series visualization charts of shopping behavior,''} said $E_C$.

\section{Discussion}\label{sec:discussion}

In this section, we summarize the implications of our work and lessons learned during our interdisciplinary collaboration with data privacy experts. We also discuss the generalizations, limitations, and future work of \textit{PriVis} and \textit{DPVisCreator}.

\textbf{Implications.} To the best of our knowledge, our work is the first to extend DP to multiple visualization chart publishing. \rc{Previous work~\cite{zhang2020investigating} has offered preliminary suggestions for preserving salient visual patterns in private visualizations. We took a step forward in proposing a Bayesian network-based DP data publishing method, taking into account user preferences for data patterns in different charts. Our approach can help data custodians make trade-offs on privacy protection and visualization utility in privacy-critical areas such as healthcare and finance.} We hope our work will move privacy-preserving visualization research forward and inspire broader cooperation between data privacy experts and visualization researchers.


\textbf{Lessons learned.} \rc{We have gained valuable experience from our interdisciplinary research in visualization and privacy. First, choosing an effective solution requires knowledge and practical experience in both domains. We initially sought DP algorithms for basic query operations, such as calculating averages and sum. However, the visualization generation process involves many interrelated data transformations, such as binning, aggregating, and sorting, which either do not have corresponding privacy-protected versions or only have theoretical derivations awaiting further experiments to illustrate their efficacy. As such, we chose a more general DP data generation approach to obtain private visualizations.}

\rc{Second, incorporating the user's prior knowledge into the privacy-preserving process can lead to more refined and customized results. The traditional automatic data-centric methods do not consider the different emphasis of users on data. In contrast, the human-centric approach allows users to analyze and compare the impact of privacy protection on charts through a visualization interface, which facilitates users to interactively guide the model until their needs are met, thus improving the utility of the results.}


\textbf{Generalizability.} The generalizability of our approach is twofold. Firstly, \textit{PriVis} is designed for tabular data. Since we use pattern constraints and original data to generate Bayesian networks, privacy-preserving charts based on tables can be generated as long as data corresponding to the chosen patterns is available. However, applying it to other data types such as graphs and geospatial data requires additional effort. \rc{These data have a particular concept of DP~\cite{kasiviswanathan2013analyzing, nie2016geospatial}. For example, graphs need to consider information that may be leaked by sub-structures such as nodes and edges, which is beyond the scope of this paper.} \rc{Secondly, for different utility metrics, we realize that the definition of utility changes with different tasks. As \textit{DPVisCreator} is developed in a modular way, users can replace the evaluation metrics according to their needs to adapt to more diverse working scenarios.}


\textbf{Limitations and future work.} Two limitations are observed in this study. First, an important concern is scalability. \rc{After the preliminary test, the computation time of \textit{PriVis} is around 2 minutes with 1,000 data records and 15 dimensions, which means that it may be difficult for users to get timely feedback.} Extending privacy protection to big data is still a challenging problem~\cite{jain2018differential}. In the future, we plan to use parallelization~\cite{madsen2017parallel} or progressive visualization~\cite{stolper2014progressive} to accelerate the learning of Bayesian networks and visualization rendering. 

\rc{Second, our approach may be too complicated to understand. Although we have some prerequisites for data custodians using \textit{DPVisCreator}, it is nontrivial for them to understand DP, the Bayesian model, and visualization design. At the same time, the current system lacks some guidance in parameter settings, which relies on the data custodians' domain knowledge. We have also collected feedback on the desire to add an automatic recommendation for privacy protection schemes. However, since obtaining an optimal privacy protection solution is an NP-hard problem~\cite{atallah1999disclosure}, we currently use a human-in-the-loop method to give users the initiative to generate solutions iteratively. In the future, we plan to add guides and annotations to the system, collect the needs of more general users and study the relationship between descriptive privacy-utility requirements and parameter settings.}



\section{Conclusion}\label{sec:conclusion}

This paper presents a novel visual analytics approach that facilitates the data custodian to publish multiple privacy-preserving visualizations and make tradeoffs between data pattern maintenance and privacy protection. We propose \textit{PriVis}, an extended Bayesian network-based DP model, and introduce pattern constraints to model user preferences for different patterns, which are then incorporated into the privacy-preserving process. We develop a visual analytics system, \textit{DPVisCreator}, with multiple interactive and coordinated views, which assists the data custodian in specifying pattern constraints, understanding the privacy model, and evaluating privacy protection schemes. The effectiveness of our approach is demonstrated through quantitative evaluations, case studies, and expert reviews.


\acknowledgments{
We would like to thank all the reviewers for their constructive comments. 
This work was supported by the National Natural Science Foundation of China (62132017 and 62106218) and Zhejiang Lab (2022NF0AC01).

}

\bibliographystyle{abbrv-doi}

\bibliography{template}

\end{document}